**THE PIXL INSTRUMENT ON THE MARS 2020 PERSEVERANCE ROVER** A.C. Allwood[1], J.A. Hurowitz[2], B. Clark[3], L. Cinquini[1], S. Davidoff[1], R.W. Denise[1], W.T., Elam[4], M.C. Foote[1], D.T. Flannery[5], J.H. Gerhard[1], J.P., Grotzinger[6], C.M. Heirwegh[1], C. Hernandez[1], R.P. Hodyss[1], M.W. Jones[5], J.L. Jørgensen[7], J. Henneke[7], P.R. Lawson[1], Y. Liu[1], H. MacDonald[2], S.M. McLennan[2], K.R., Moore[1], M. Nachon[8], P. Nemere[5], L. O'Neil[4], D.A.K. Pedersen[7], K.P. Sinclair[4], M.E. Sondheim[1], E. Song[1], N.R. Tallarida[1], M.M. Tice[8], A. Treiman[9], K. Uckert[1], L.A. Wade[1], J.D. Young[1], P. Zamani[1], [1]NASA-Jet Propulsion Laboratory, 4800 Oak Grove Drive, Pasadena, CA, [2]Department of Geosciences, Stony Brook University, Stony Brook, NY, joel.hurowitz@stonybrook.edu, [3]Space Science Institute, Boulder, CO, [4]University of Washington, Seattle, WA, [5]Queensland University of Technology, Brisbane, AU, [6]California Institute of Technology, Pasadena, CA, [7]Danish Technical University, Lyngby, DK, [8]College of Geosciences, Texas A&M University, College Station, TX, [9]Lunar and Planetary Institute, Houston, TX

**Introduction:** The Planetary Instrument for X-ray Lithochemistry (PIXL) is a micro-focus X-ray fluorescence spectrometer mounted on the robotic arm of NASA's *Perseverance* rover [1]. PIXL will acquire high spatial resolution observations of rock and soil chemistry, rapidly analyzing the elemental chemistry of a target surface. In 10 seconds, PIXL can use its powerful 120 µm diameter X-ray beam to analyze a single, sand-sized grain with enough sensitivity to detect major and minor rock-forming elements, as well as many trace elements. Over a period of several hours, PIXL can autonomously scan an area of the rock surface and acquire a hyperspectral map comprised of several thousand individual measured points. When correlated to false-color images [1, 2] acquired by PIXL's Micro-Context Camera (MCC), built by the Danish Technical University, these maps reveal the distribution and abundance variations of chemical elements making up the rock, tied accurately to the physical texture and structure of the rock, at a scale comparable to a 10X magnifying geological hand lens. The 100's to 1000's of spectra derived from PIXL scans may be analyzed individually or summed together to create a bulk rock analysis, or subsets of spectra may be summed, quantified, analyzed, and compared using PIXLISE [1, 3], a data visualization software package developed by the PIXL team in collaboration with the Queensland University of Technology. This hand lens-scale view of the petrology and geochemistry of materials at the *Perseverance* landing site will provide a valuable link between the larger, centimeter- to meter-scale observations by Mastcam-Z, RIMFAX, and Supercam, and the much smaller-scale measurements that will be made on returned samples in terrestrial laboratories.

**PIXL Science Goals:** The PIXL science investigation has three goals: 1) To provide detailed geochemical assessment of past environments, habitability, and biosignature preservation potential; 2) To detect any potential chemical biosignatures that are encountered and characterize the geochemistry of any other types of potential biosignatures detected; 3) To provide a detailed geochemical basis for selection of a compelling set of samples for return to Earth.

Achieving these goals requires careful, integrated observation of key rock properties at regional to outcrop to micro scales: one of the key properties to measure is fine-scale elemental chemistry. Specifically, submillimeter-scale abundance and distribution of elements is required because the origin, biogenicity, and paleo-environmental significance of the elements is strongly dependent on their distribution relative to individual rock components. These components are commonly present at submillimeter scales and can represent vastly different phases of the rock's history, from pre-genetic (detrital grains, clasts) to syngenetic (precipitated laminae, authigenic grains and cements) to post-genetic (cross-cutting veins, weathering fronts).

**Instrument Overview:** PIXL consists of three major hardware components:

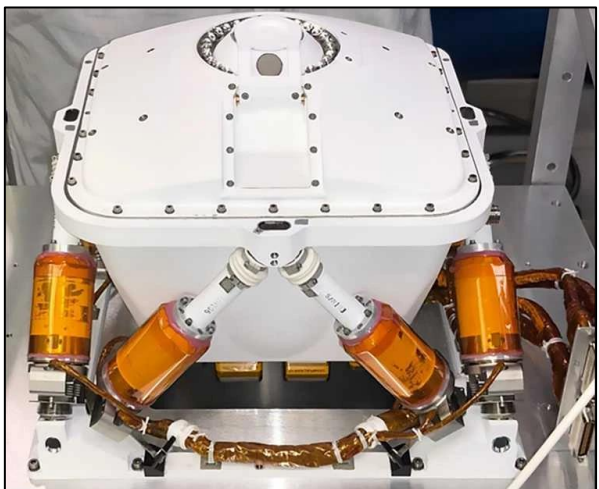

**Fig. 1**: The PIXL sensor assembly prior to integration on the rover arm turret.

1. The sensor assembly (**Fig. 1**), mounted on the instrument turret at the end of the rover arm, contains the key instrument hardware that generates, focuses, and detects X-rays, takes images with flash illumination, and measures distance to target by imaging arrays of laser dots projected by Structured Light Illuminators (SLI's) onto the target. The sensor assembly is connected to the rover-arm turret via six actuated struts arranged in three pairs. These struts combine to adjust the position of the sensor head in X, Y, Z, tilt, and tip, which enables pointing of the sensor head at a specified location, adjustment of focus, and scanning.

2. The body unit electronics (BUE), mounted inside the rover body, contains most of the instrument's electronic controls, including a separate MCC electronics unit. The BUE is connected electrically to the sensor assembly by approximately 12 meters of a combination of flex and round wire harnessing.

3. The calibration target (**Fig. 2**) is mounted on the rover arm azimuth motor housing. Features of the calibration target include discs of PTFE, BHVO-2 glass, NIST-610 glass, the mineral scapolite (mounted in epoxy), a glass disk with a metal cross composed of a chromium line in one axis and nickel line in the other axis, and a pseudo-random distribution of black dots, divided between two surfaces separated by 5 mm in depth. These features enable calibration of the X-ray subsystem and of the geometric alignment of the MCC, SLIs (collectively called the "Optical Fiducial Subsystem, or OFS"), and the X-ray beam.

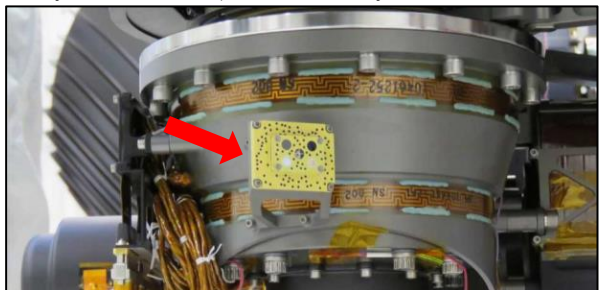

**Fig. 2**: Location of the PIXL calibration target (red arrow) on the rover arm azimuth motor housing.

**PIXL Scans:** PIXL measurements are grouped into three basic types: line, grid, and map scans. An example of a line-scan activity is shown on **Fig. 3**, which also shows some of the activities that take place during a PIXL scan. These activities include: 1) XRF analyses; 2) "OFS checks", which assess instrument standoff from the target to determine if the X-ray beam needs to be refocused; 3) "Terrain Relative Translation (TRT) checks", in which the OFS determines the offset, caused by arm or rover thermal drift, between the actual and planned pointing location in the scan. TRT checks inform corrective pointing of the sensor head. 4) "MCC context" images of the target surface. For this example, where the step size between XRF analysis locations is set equal to the beam diameter (120 μm), and the XRF dwell duration at each step is set to 10 seconds, the scan would require approximately 3 hours to complete (~1.6 hour of X-ray integration on 600 points, with the remaining time devoted to imaging, scan motions, and

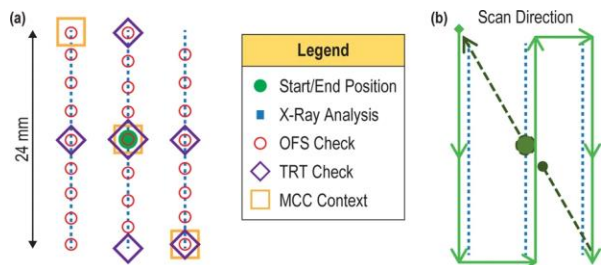

**Fig. 3**: Line scan activity consisting of 3 columns spaced 5 mm apart. (a) Activities that take place during a scan (see legend and text). (b) Direction of travel in the scan.

X, Y, Z position checking). This scan would generate ~25 Mbits of data volume, of which 10.5 Mbit would be earmarked for immediate downlink to seed the generation of decisional data products.

Grid scans involve scanning the X-ray beam in two dimensions across a target in step sizes that are significantly larger than the X-ray beam diameter. Maps involve scanning the X-ray beam in two dimensions across a target in step sizes equal to the X-ray beam diameter. Maps are the highest-density, most information-rich measurements, but they also require significantly more time to execute (up to ~16 hours). Accordingly, they are conducted at night, when the rover is asleep and thermal drift is minimized.

**PIXL Data and PIXLISE:** PIXL data consist mainly of X-ray spectra, images, spatial information collected by the OFS that allows reconstruction of target topography and X-ray beam location, and instrument housekeeping data. After downlink, these data are processed by the PIXL instrument Science Data System [1], producing science and engineering "Auto- and Quicklook" data products for rapid decision making on the tactical timeline. In-depth analysis of PIXL data is conducted with PIXLISE [1, 3], cloud-based visual analytics [4] tool that allows scientists to see multiple visual mappings of the data at a single time. Quantification of PIXL XRF data can be performed from within PIXLISE using PIQUANT [1], an in-house fundamental parameter physics-based code that incorporates X-ray parameter databases [5].

**Expected Results:** By the time of LPSC 2021, *Perseverance* will have completed its 30-sol "Surface Operations Transition Phase" (called SOX). We will provide an instrument overview, a report on our first in-flight calibration activity during SOX, and any opportunistic science activities that have occurred.

**References:** [1] Allwood A.C., et al. (2020) *SSR, 216,* Article #134. [2] Liu Y. et al. (2021) *LPSC 52*. [3] Schurman, D. et al. (2019) *Proc. AbSciCon 2019*. [4] Keim et al. (200) *in* Visual Data Mining, S.J. Simoff, et al., eds. Lecture Notes in Computer Science, *4404*. [5] Elam, W.T., et al. (2002) *Rad. Phys. Chem.* 63, 121-128.